\documentclass[amsmath,amssymb,pra,aps,showpacs,twocolumn,10pt]{revtex4-1}
\usepackage{amsmath,amsfonts,amssymb,amsthm,graphics,graphicx,epsfig,bbm}
\usepackage[colorlinks=true,citecolor=blue,linkcolor=red,urlcolor=blue]{hyperref}
\usepackage[usenames]{color}
\usepackage{graphicx}
\usepackage{subfigure}
\usepackage{amsmath}
\usepackage{epsfig}
\usepackage{dcolumn}
\usepackage{bm}
\usepackage{color}
\usepackage{epstopdf}
\usepackage{amssymb}
\usepackage{amstext}
\usepackage{latexsym}
\usepackage{hyperref}
\usepackage{amsfonts}
\usepackage{psfrag}
\usepackage{xcolor}
\usepackage[normalem]{ulem}
\usepackage{dsfont}
\usepackage{txfonts}
\usepackage{overpic}
\usepackage{ifthen}
\usepackage{amsthm}

\usepackage{amsmath, amssymb, amsthm}

\newcommand{\ket}[1]{\vert #1 \rangle}
\newcommand{\bra}[1]{\langle #1 \vert}

\newcommand{\an}[2]{\ifthenelse{\equal{#1}{}}{\ensuremath{\hat{#1}_{#2}}}{\ensuremath{\hat{#1}^{\protect\phantom{\dagger}}_{#2}}}}

\newtheorem*{remark}{Remark}

\begin{document}

\title{Quantum Signal Processing with the one-dimensional quantum Ising model}

\author{V. M. Bastidas$^1$}
\email{victor.bastidas@ntt.com}
\altaffiliation[]{These authors contributed equally to this work.}
\author{S. Zeytino\u{g}lu$^{2,3}$}
\altaffiliation[]{These authors contributed equally to this work.}
\author{Z. M. Rossi$^{4,2,1}$}
\altaffiliation[]{These authors contributed equally to this work.}
\author{I. L. Chuang$^{4}$}
\author{W. J. Munro$^{1}$}
\affiliation{%
$^1$NTT Basic Research Laboratories \& Research Center for Theoretical Quantum Physics,  3-1 Morinosato-Wakamiya, Atsugi, Kanagawa, 243-0198, Japan} 
\affiliation{%
$^2$Physics and Informatics Laboratory, NTT Research, Inc., 940 Stewart Dr., Sunnyvale, California, 94085, USA
} 
\affiliation{%
$^3$Department of Physics, Harvard University, Cambridge, Massachusetts 02138, USA
} 
\affiliation{%
$^4$Department of Physics, Massachusetts Institute of Technology, Cambridge, Massachusetts 02139, USA
} 
\date{\today}

\begin{abstract}

Quantum Signal Processing (QSP) has emerged as a promising framework to manipulate and determine properties of quantum systems. QSP not only unifies most existing quantum algorithms but also provides tools to discover new ones. Quantum signal processing is applicable to single- or multi-qubit systems that can be “qubitized” so one can exploit the SU$(2)$ structure of system evolution within special invariant two-dimensional subspaces. In the context of quantum algorithms, this SU$(2)$ structure is artificially imposed on the system through highly nonlocal evolution operators that are difficult to implement on near-term quantum devices. In this work, we propose QSP protocols for the infinite-dimensional Onsager Lie Algebra, which is relevant to the physical dynamics of quantum devices that can simulate the transverse field Ising model. To this end, we consider QSP sequences in the Heisenberg picture, allowing us to exploit the emergent SU$(2)$ structure in momentum space and ``synthesize” QSP sequences for the Onsager algebra. 
Our results demonstrate a concrete connection between QSP techniques and Noisy Intermediate Scale quantum protocols. We provide examples and applications of our approach in diverse fields ranging from  space-time dual quantum circuits and quantum simulation, to quantum control.

\end{abstract}

\maketitle
\section{Introduction}
Originally inspired by composite pulse sequences in nuclear magnetic resonance (NMR), quantum signal processing (QSP) has emerged as a framework to unify existing quantum algorithms and discover new ones using well-developed tools from functional analysis~\cite{Low2017,Martyn2021,Rossi2022,Rossi2023semantic}. QSP is a successful framework for precisely controlling the evolution of quantum systems when one is given repeatable access to basic quantum processes (unitary evolutions). The iterative structure of QSP appears in many contexts, and suggests the applicability of similar ideas to improve understanding of control protocols in many-body quantum systems. Indeed, most explorations into the non-equilibrium behavior of condensed matter systems~\cite{Polkovnikov2011,Hatomura2022}, including those studying quantum annealing~\cite{Das2008,Barends2016,Mbeng2019,Hatomura2018,Palacios2022,Bastidas2022,Hatomura2023}, discrete time crystals~\cite{Sacha2018,Else2020,Estarellas2020,Sakurai2021}, and space-time dual quantum circuits~\cite{Akila2016,Piroli2020,Bertini2018,Lu2021,Fisher2023}, rely on the fact that the dynamics depend on iterated processes. If this structural similarity is sufficient to import QSP techniques and precisely control many-body quantum systems currently realized in experiments~\cite{Zeytinouglu2022}, we can expand both our understanding of non-equilibrium dynamics and our capacity to control and manipulate the quantum systems.

The application of QSP protocols in current experimental platforms is difficult as conventional circuit instantiations of QSP protocols rely on highly nonlocal unitaries that are difficult to implement in Noisy-intermediate scale quantum (NISQ) devices. QSP and its multi-qubit extension, quantum singular value transformation (QSVT)~\cite{Gilyen2019}, rely on strong
conditions known as qubitization~\cite{Low2019hamiltonian,Martyn2021}, which ensure that the dynamics of the system can be described as a direct sum of two-dimensional subspaces whose dynamics are summarizable in terms of SU$(2)$ operations. The two conventional methods to impose such a structure rely on the use of highly non-local interactions~\cite{Martyn2023}. In the first method, qubitization~\cite{Low2019hamiltonian,Martyn2021} can be imposed on the dynamics by implementing a highly non-local partial reflection operation acting on the whole system. In the second method, one uses non-local interactions between the system and a single ancilla to condition the dynamics of the system on the ancilla. Then, the tensor product structure can be used to endow the overall dynamics with the desired behavior. More recently, Refs.~\cite{Lloyd2021,Zeytinouglu2022} propose more natural implementations of QSP protocols. However, these restricted protocols still rely on highly non-local interactions between the system and a single ancillary qubit. Hence, whether the qubitization conditions can be satisfied for the dynamics of an extended system evolving under local dynamics will determine the applicability of QSP to the study of near-term many-body quantum systems. Moreover, there are mathematical challenges in trying to use QSP for multiqubits systems that are not qubitized and for other Lie groups beyond SU(2). A recent effort in this direction is the development QSP algorithms for continuous variables described by the SU(1,1) Lie group~\cite{rossi2023}. Most importantly, QSP is a framework built in the context of finite dimensional vector spaces. Consequently the validity of applying similar techniques to the analysis of infinite dimensional systems is not obvious. We show below that this condition requires us to either simplify how we represent these systems (by identifying underlying symmetries) or substantially alter the basic structure of QSP.

In this paper we apply QSP-inspired techniques to the one-dimensional quantum transverse-field Ising model (TFIM), a condensed matter system which is of general theoretical interest from quantum annealing~\cite{Das2008} to space-time dual circuits~\cite{Akila2016,Lu2021}, and one that is routinely realized experimentally in diverse NISQ platforms~\cite{mi2022time,Mi2022Resilient}. We define QSP sequences for the Onsager algebra~\cite{Onsager1944,Davies1990,Uglov1996,Gritsev2017}. This is an infinite-dimensional Lie algebra underlying the solution of the Ising model that shares some basic traits with the su$(2)$ algebra undergirding conventional QSP. We further determine the conditions under which repeated access to unitary evolutions induced by Hamiltonian terms of the TFIM allows one to implement generic QSP protocols. Lastly, we highlight the power of the proposed QSP sequences by applying them to a wide range of scenarios of current interest, ranging from space-time dual quantum circuits and Hamiltonian engineering to composite pulse sequences in spin systems.

To achieve these results, we rely on two core ingredients. First, we use a Jordan-Wigner mapping~\cite{jordan1928} between the TFIM and a non-interacting fermionic model~\cite{sachdev_2011}. Because TFIM is integrable, the associated fermionic Hamiltonian is a quadratic form in terms of fermionic ladder operators in each momentum sector. Second, unlike the conventional QSP approach that is interested in state evolution, we consider the action of the Hamiltonian evolution operator on the fermionic ladder operators in the Heisenberg picture. The terms in the TFIM Hamiltonian generate SU$(2)$-like transformations of fermionic operators. These transformations are then cascaded into QSP-like iterative protocols, defined by a set of parameters each assigned for one iteration. We then identify the special points in the parameter space for which the evolution is as expressible as standard QSP acting on the space of fermionic operators.

QSP and its related algorithms are far more flexible than initially considered. Under specially tuned conditions, the evolution of many complex condensed matter systems is succinctly described and controlled by methods that are quite similar to QSP, even when they are evolving under local dynamics. The QSP methodology brings new insights into our understanding of the dynamics of quantum systems and allows us to design novel control sequences to improve the performance of near-term quantum devices. We discuss the application of QSP methods to dual quantum circuits~\cite{Akila2016,Piroli2020,Bertini2018,Lu2021,Fisher2023}, which could be used to define QSP sequences in hybrid quantum circuits composed of unitary operations and measurements\cite{Lu2021}. Moreover, we show that the proposed QSP sequences can be used to control the dynamics of single-particle fermionic excitations by engineering their dispersion relation. Similarly, we can use this ability to engineer the single-particle dispersion relation to simulate various spin Hamiltonians which correspond to non-interacting fermionic Hamiltonians.

Our results point towards further challenges for QSP to subsume, as well as avenues toward the utility of QSP protocols in describing locally interacting multi-qubit systems. Unlike in the standard case, QSP in the Heisenberg picture can be easily extended to the non-unitary evolution of the fermionic operators by using the space-time duality~\cite{Akila2016,Piroli2020,Bertini2018,Lu2021,Fisher2023}.  Additionally, QSP-like sequences of SU$(2)$ transformations will allow us to design control sequences for a wider range of experimental scenarios and to strengthen our understanding of iteratively-evolved quantum mechanical systems.

The structure of our paper is as follows. In section~\ref{SecII} we provide a brief summary of conventional QSP using SU$(2)$ operations. In section~\ref{SecIII} we introduce the Onsager Lie algebra and Krammers-Wannier duality, and define the QSP sequences terms of the ``seed operators" of this Lie algebra. In section~\ref{SecIV} we discuss the intimate relation between Onsager algebra and the Ising model and discuss the physical implementation of QSP in terms of single- and two-qubit operations. In section~\ref{SecV}  we demonstrate that after a Jordan-Wigner transformation, we can obtain simple QSP sequences for fermionic operators in the Heisenberg picture when we work in momentum space. We also discuss the expressivity of QSP in the Heisenberg picture. In addition, in section~\ref{SecVI} we provide specific examples of QSP sequences using Onsager algebra in the context of space-time dual quantum systems, Hamiltonian engineering and composite pulse sequences in spin chains. Lastly, we
provide concluding remarks and an outlook in section~\ref{SecVII}.

\section{Quantum signal processing (QSP) revisited \label{SecII}}
In nuclear magnetic resonance (NMR) there exist many composite pulse techniques designed to achieve specific goals, such as the precise control of the dynamics of quantum systems~\cite{Wimperis1994,minch1998spin,Vandersypen2005,Mount2015,Low2016} and the reduction of noise. One can think of a sequence of parameterized unitary operations, in analogy to how they are used in NMR, as a means to calculate a response function. Recently, Quantum Signal processing (QSP) has emerged as general theory of composite pulse sequences, and has proven itself as a versatile approach to design quantum circuits, ultimately permitting the unification and simplification of most of the known quantum algorithms~\cite{Low2017,Martyn2021}. In the language of QSP, a sequence of unitaries allows one to process an unknown signal encoded in said unitaries, such that measurement results can depend on said signal in highly-non-linear, near arbitrary ways~\cite{Martyn2021}. 

We briefly summarize the major takeaways of QSP in terms of the su$(2)$ algebra by first defining the signal operator~\cite{Martyn2021}
\begin{align}
          \label{eq:SignalOperatorQubit}
\hat{W}(x)=e^{\mathrm{i} \frac{\delta}{2}X}=\begin{bmatrix}
    x &  \mathrm{i} \sqrt{1-x^2}  \\
     \mathrm{i}\sqrt{1-x^2}  & x  \\
    \end{bmatrix}
\ ,
\end{align}
where $\delta=-2\cos^{-1}x$ with $x\in [-1,1]$ while ${X,Y,Z}$ are Pauli matrices generating the su$(2)$ Lie algebra. The signal $\delta$ is processed through a sequence of rotations that do not commute with the signal operator, defined by
\begin{align}
          \label{eq:SignalProcessingOperatorQubit}
\hat{S}(\phi_l)=e^{\mathrm{i} \phi_l Z}
\ .
\end{align}
If the sequence contains $d+1$ rotations used to process the signal, it is convenient to organize the angles into a vector $\vec{\phi}=(\phi_0,\phi_1,\dots\phi_d)$. A theorem of QSP establishes that given QSP sequence parameterization $\hat{U}_{\vec{\phi}}$ induces a polynomial transformation of $x$ as follows
\begin{align}
\hat{U}_{\vec{\phi}}&=e^{\mathrm{i} \phi_0 Z}\prod^d_{r=1} \hat{W}(x) e^{\mathrm{i} \phi_r Z}=\begin{bmatrix}
    P(x) &  \mathrm{i} Q(x) \sqrt{1-x^2}  \\
     \mathrm{i}Q^*(x) \sqrt{1-x^2}  & P^*(x)  \\
    \end{bmatrix}
\ .\label{eq:SignalProcessingOperatorQubit}
\end{align}
Further, there is a sequence of rotations $\vec{\phi}$ for any polynomials $P(x)$ and $Q(x)$ satisfying mild requirements~\cite{Martyn2021} on parity and norm. The cornerstone of this result is that, given function one wishes to apply to $x$, there exists an efficiently computable sequence of angles $\vec{\phi}$ encoding a polynomial approximation of it~\cite{Low2016}.

In its original form~\cite{Martyn2021}, this theorem was defined considering the structure of the $\text{su}(2)$ algebra $[X,Y]=2\mathrm{i}Z, \ [Y,Z]=2\mathrm{i}X, \text{and} \ [Z,X]=2\mathrm{i}Y$ which is finite dimensional and generates both the signal and signal processing operators belonging to  the compact group SU$(2)$. The simple form of the QSP operation sequence $\hat{U}_{\vec{\phi}}$ is possible due to well-known properties of the Pauli matrices, e.g., $X^2=Y^2=Z^2=\hat{1}$.
Previous works mostly use qubitization to obtain QSP sequences in a many-qubit system by exploiting the SU$(2)$ dynamics within two-dimensional invariant subspaces~\cite{Low2019hamiltonian,Martyn2021}. However, this procedure either requires controlled versions of $n$-qubit unitaries (requiring extra ancillae) or evolutions generated by $n$-qubit reflection operators (which have to be highly nonlocal). Hence, it is desirable to find QSP-like schemes that are easy to implement with local interactions. 

The question we want to answer in this work is whether QSP sequences can be defined in infinite dimensional Lie algebras~\cite{kac1990infinite} such as the Kac-Moody~\cite{Dolan1981,wan1991introduction} or the Virasoro algebra in conformal field theory~\cite{Friedan1984,francesco2012conformal}. These algebras play an important role in diverse fields ranging from low-energy regimes (low temperatures and long wavelength excitations) in condensed matter physics~\cite{Orgad1997,von1998bosonization} to high-energy physics~\cite{goddard1986kac} and string theory~\cite{Schwarz1999}. 

For concreteness, in this work, we focus on the Onsager algebra appearing in the Ising model, which is an infinite-dimensional algebra of importance in statistical physics and the study of critical phenomena~\cite{Onsager1944,Davies1990,Uglov1996,Gritsev2017}. For instance, this algebra has representations as transfer matrices of the classical 2D Ising model~\cite{Onsager1944}. In the next section, we briefly summarize the basic aspects of the Onsager algebra and provide its representation in terms of the quantum Ising model~\cite{sachdev_2011}.

\section{QSP with the Onsager Lie algebra\label{SecIII}} 
In the previous section, we discussed how QSP depends on the su$(2)$ algebra. In this section, we explore an infinite-dimensional algebra known as the Onsager algebra~\cite{Onsager1944,Uglov1996,Gritsev2017}, widely used in statistical physics and theory of integrability~\cite{Uglov1996,Gritsev2017}. The Onsager algebra is defined in terms of operators $\hat{A}_n$ and $\hat{G}_n$, which are recursively generated from ``seed" operators $\hat{A}_0$ and $\hat{A}_1$ via the following relations
\begin{align}
          \label{eq:SeedOnsagerAlgebra}
[\hat{A}_n,\hat{A}_0]&=4G_n
\nonumber\\
[\hat{G}_1,\hat{A}_n]&=2(\hat{A}_{n+1}-\hat{A}_{n-1})
\ .
\end{align}
From these relations it is possible to build the complete structure of the algebra, as follows
\begin{align}
          \label{eq:SeedOnsagerAlgebra}
[\hat{A}_n,\hat{A}_m]&=4\hat{G}_{n-m}
\nonumber\\
[\hat{G}_n,\hat{A}_m]&=2(\hat{A}_{m+n}-\hat{A}_{m-n})
\nonumber\\
[\hat{G}_n,\hat{G}_m]&=0
\ ,
\end{align}
where $\hat{G}_{-n}=-\hat{G}_n$. An important aspect of this algebra is that the ``seed" operators should satisfy the so called Dolan-Grady conditions~\cite{Gritsev2017}
\begin{align}
          \label{eq:SeedOnsagerAlgebra}
[\hat{A}_0,[\hat{A}_0,[\hat{A}_0,\hat{A}_1]]]&=16[\hat{A}_0,\hat{A}_1]
\nonumber\\
[\hat{A}_1,[\hat{A}_1,[\hat{A}_1,\hat{A}_0]]]&=16[\hat{A}_1,\hat{A}_0]
\ .
\end{align}
These relations reveal a fundamental symmetry of statistical mechanics known as the Krammers-Wannier duality~\cite{Kramers1941,Kogut1979,2017Motrunich}, which is related to the theory of the two-dimensional Ising model and the one-dimensional quantum Ising model in a transverse field. More specifically, the duality means that we can get an equivalent theory by exchanging the ``seeds" of the algebra as follows: $\hat{A}_0\rightarrow \hat{A}_1$ and $\hat{A}_1\rightarrow \hat{A}_0$. 

Before discussing any particular representation of the Onsager algebra, let us explore the feasibility of defining a QSP sequence using the generators $\hat{A}_n$, as they are the fundamental units used to build the full algebra. As the algebra is constructed in a recursive fashion, it is reasonable to define a QSP using the exponential map $\exp:\mathcal{G}\rightarrow G$, allowing one to map a Lie algebra $\mathcal{G}$ to a corresponding Lie group~\cite{fegan1991introduction}. From now on, we will assume the existence of an infinite-dimensional unitary representation of the group $G$ associated to the Onsager algebra.

Inspired by the definition of QSP in the case of a single qubit, we define here the signal operator
\begin{align}
          \label{eq:OnsagerSignalOperator}
\hat{W}^{\text{O}}(\theta)=\exp\left(\mathrm{i} \theta \hat{A}_1\right)
\ .
\end{align}
Correspondingly, let us also define the signal-processing unitary operator
\begin{align}
   \label{eq:OnsagerSignalProcessingOperator}
\hat{S}^{\text{O}}(\phi_r)=\exp\left(\mathrm{i}\phi_r \hat{A}_0\right)
\ .
\end{align}
\begin{figure}
	\includegraphics[width=0.48\textwidth]{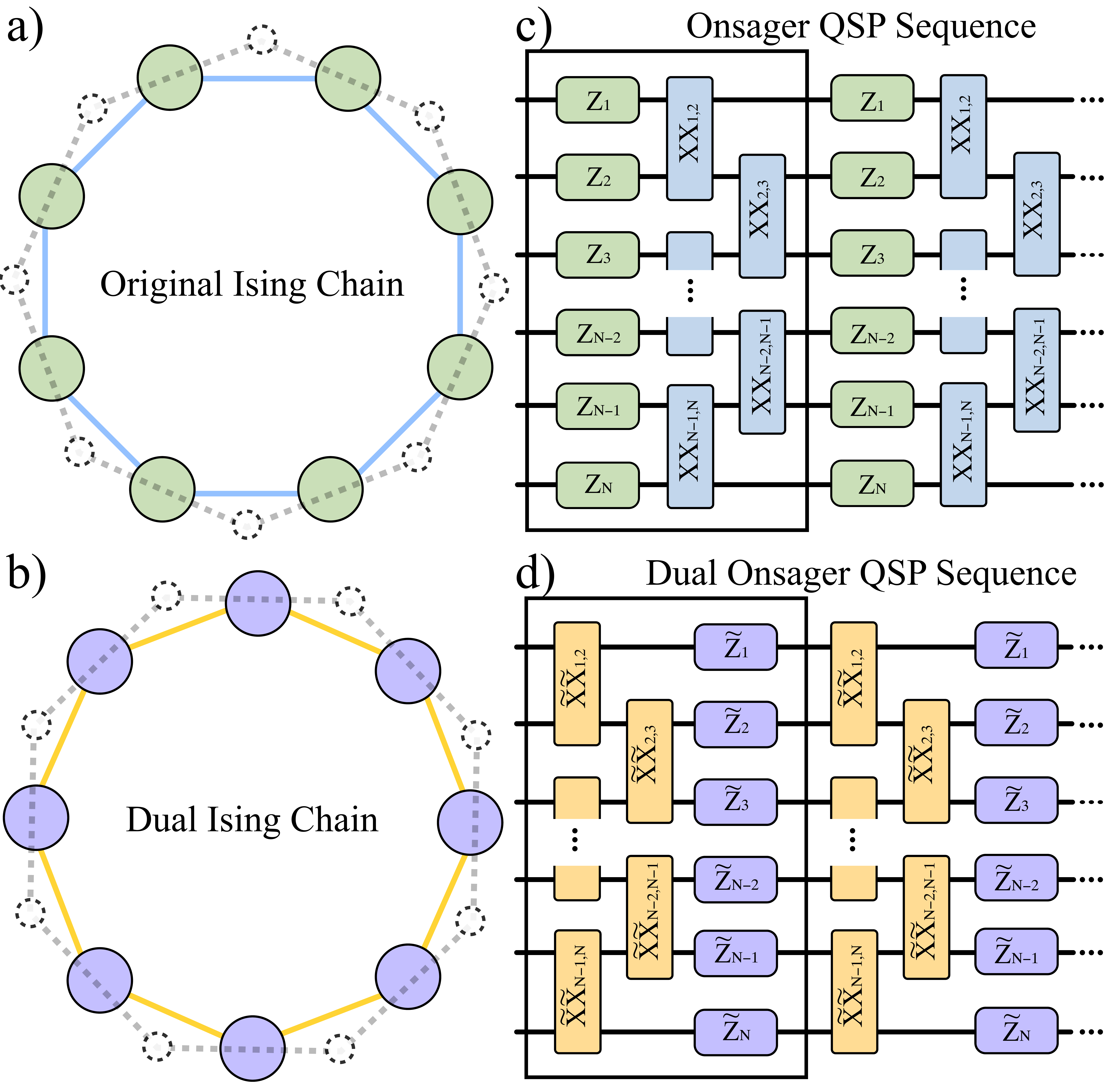}
	\caption{QSP with the Ising chain and Krammers-Wannier duality. Here a) and b) illustrate the Ising chain and its dual. Moreover, c) and d) depicts the corresponding quantum circuit to implement the QSP sequence dependent on the Onsager algebra. Under the duality transformation, lattice sites map to links in the dual lattice and vice-versa. The dashed lines in a) and b) show the links and sites of the dual and original lattice, respectively. In terms of a practical implementation of our ideas in NISQ devices, the lattice sites and bonds in a) and b) represent the single- and two-qubit gates in c) and d), respectively. }
	\label{Fig1}
\end{figure}
Considering the combined action of these two operators, we can define a QSP variant in terms of Onsager generators:
\begin{align}
          \label{eq:OnsagerQSP}
\hat{U}^{\text{O}}_{\vec{\phi}}(\theta)=\hat{S}^{\text{O}}(\phi_0)\prod^d_{r=1}\hat{W}^{\text{O}}(\theta)\hat{S}^{\text{O}}(\phi_r)
\ .
\end{align}
Furthermore, in contrast to previous works in QSP, the Dolan-Grady conditions~\cite{Davies1990,Uglov1996,Gritsev2017} allow us to build a ``Dual Onsager QSP" sequence
\begin{align}
          \label{eq:DualOnsagerQSP}
\hat{U}^{\text{DO}}_{\vec{\phi}}(\theta)=\hat{S}^{\text{DO}}(\phi_0)\prod^d_{r=1}\hat{W}^{\text{DO}}(\theta)\hat{S}^{\text{DO}}(\phi_r)
\ .
\end{align}
by exchanging $\hat{A}_0\rightarrow \hat{A}_1$ and $\hat{A}_1\rightarrow \hat{A}_0$ in Eq.~\ref{eq:OnsagerQSP}. Here, $\hat{W}^{\text{DO}}(\theta)$ and $\hat{S}^{\text{DO}}(\phi_r)$ are the dual signal and signal processing operators.

Although the nature of the  Onsanger algebra is fundamentally different from that of su$(2)$ in standard QSP, this modified QSP sequence still exploits the non-commuting character of the ``seed" operators to build up a nontrivial set of physical operations. We can consider a
spin representation of the Onsager algebra~\cite{Gritsev2017} with ``seed" operators $\hat{A}_0=\sum^{N}_{j=1}  X_j$ and $\hat{A}_1=\sum^N_{j=1}Z_j Z_{j+1}$ where $X_j, Y_j, Z_j$ are Pauli matrices at a given site $j$ with periodic boundary conditions $X_{N+1}=X_1,Y_{N+1}=Y_1, \text{and} \ Z_{N+1}=Z_1$. In this case, we can explicitly see the nontrivial character of the duality that maps product states (eigenstates of $\sum^{N}_{j=1}  X_j$) to maximally-entangled states (eigenstates of $\hat{A}_1=\sum^N_{j=1}Z_j Z_{j+1}$). 
\section{The Onsager Lie algebra and the one dimensional quantum Ising model in a transverse field \label{SecIV} } 
To implement a manybody version of quantum signal processing, one needs to build a discrete sequence of physical operations that can be interpreted as a program to calculate a desired function. As such to build a discrete sequence of operations in a manybody system, we focus here on a time dependent one dimensional quantum Ising model~\cite{sachdev_2011}
\begin{align}
          \label{eq:IsingHamitonian}
\hat{H}(t)=-\hbar g(t)\sum^{N}_{j=1} Z_j-\hbar J(t)\sum^{N}_{j=1}X_j X_{j+1}
\ ,
\end{align}
where $g(t)$ is a global time dependent transverse field while $J(t)$ is a time-dependent interaction strength. At this stage is important to emphasize that our approach requires some knowledge of $g(t)$ and $J(t)$ and in terms of a particular implementation, it requires controllability of these parameters. Recent experiments~\cite{mi2022time,Mi2022Resilient} demonstrate the high degree of control of the parameters $g(t)$ and $J(t)$ using arrays of superconducting qubits. In this way, we can build discrete single- and two-qubit operations by modulating the parameters $g_j(t)$ and $J(t)$, respectively.

The crucial point of the theory of the quantum Ising model is that the Hamiltonian Eq.~\eqref{eq:IsingHamitonian} is an integrable model built in terms of generators of the Onsager algebra ~\cite{Gritsev2017}.

With these elements at hand, we can define a manybody QSP sequence as follows
\begin{align}
          \label{eq:IsingQSP}
\hat{U}_{\vec{\phi}}(\theta)=e^{\mathrm{i}\phi_{0}\sum^{N}_{j=1} Z_j}\prod^d_{r=1}e^{\mathrm{i} \theta\sum^{N}_{j=1}X_j X_{j+1}}
e^{\mathrm{i}\phi_{r}\sum^{N}_{j=1} Z_j}
\ .
\end{align}
Now that we have establish the relation between the Ising model and the Onsager algebra, we can explore the physical meaning of the duality and understand its nontrivial character. To do this, let us consider the time independent case $g(t)=g_0$ and $J(t)=J_0$. When the transverse field strength is much stronger than the spin interaction, the system is in the paramagnetic phase. In the opposite regime, the system is in the ferromagnetic phase, which is characterized by long-range correlations between the spins. What the Krammers-Wannier duality does is to exchange the role of the terms giving us the dual Hamiltonian~\cite{2017Motrunich,Haug2020}
\begin{align}
          \label{eq:IsingHamitonianDual}
\hat{H}^{\text{D}}(t)=-\hbar g(t)\sum^{N}_{j=1} \tilde{X}_j \tilde{X}_{j+1}-\hbar J(t)\sum^{N}_{j=1}\tilde{Z}_j
\ ,
\end{align}
where $\tilde{X}_j, \tilde{Y}_j, \tilde{Z}_j$ are Pauli matrices in the dual lattice. Geometrically, this duality can be understood as replacing links by nodes and nodes by links in the chain~\cite{2017Motrunich}. At the critical point $g_0=J_0$, the system is self-dual, as the Hamiltonian looks the same both in the original and dual representations. This is not only a mathematical curiosity. In fact, as stated before, the duality can be interpreted as symmetry in statistical mechanics where the self-dual point is a quantum critical point of the model~\cite{sachdev_2011}. Further, the quantum Ising chain can be mapped to the two-dimensional classical Ising chain. The quantum critical point naturally maps to the critical temperature at which the classical phase transition occurs in the classical 2D Ising model~\cite{Onsager1944,sachdev_2011,Kramers1941}.

Next, by using the Krammers-Wannier duality, we can define the dual QSP sequence that exchanges the role of signal and signal processing operators
\begin{align}
          \label{eq:IsingDualQSP}
\hat{U}^{\text{D}}_{\vec{\phi}}(\theta)=e^{\mathrm{i}\phi_{0}\sum^{N}_{j=1}\tilde{X}_j \tilde{X}_{j+1}}
\prod^d_{r=1}e^{\mathrm{i} \theta\sum^{N}_{j=1} \tilde{Z}_j}
e^{\mathrm{i}\phi_{r}\sum^{N}_{j=1}\tilde{X}_j \tilde{X}_{j+1}}
\ .
\end{align}
Although this expression looks fairly simple, it is highly nontrivial, due to the non-commuting character of the signal and signal-processing operators. Moreover, there is a operational relation between the original and dual quantum circuits depicted in Fig.~\ref{Fig1}~c)~and~d), which is given by
\begin{align}
          \label{eq:OperationalDuality}
\hat{U}^{\text{D}}_{\vec{\phi}}(\theta)=\hat{U}^{\dagger}_{-\vec{\phi}}(-\theta)
 \ .
\end{align}
 
Next, it is important to discuss the experimental feasibility of our proposal. A recent experiment~\cite{mi2022time} implemented a spin-spin interaction of the form $\theta\sum_j Z_jZ_j$ term and the transverse field $\phi\sum_j X_j$. In their experiment, they chose values of the parameters such that $\theta\in[0.5\pi,1.5\pi]$ and $\phi\in[-\pi,\pi]$. Our model can be exactly mapped to the model realized experimentally by using spin rotations. 

Another important point that we want to emphasize is that so far, as the Onsager algebra is infinite-dimensional, the algebraic structure of the problem is not related to the su$(2)$ algebra used in the case of the single-qubit QSP. In the next section, we extend the notion of QSP sequence at the level of the operators and then map the system to a fermionic representation. This allows us to simplify the complexity of the problem.
 
\section{Jordan-Wigner transformation and QSP in the Heisenberg picture\label{SecV}} 
In this section, we briefly summarize how to use tools from the theory of the TFIM to effectively reduce the dynamics of the model to a pseudo-spin representation in the Heisenberg picture. This will enable us to work using the su$(2)$ algebra.

One of the most interesting aspects of the one-dimensional quantum Ising model is that it can be mapped to a system of non-interacting fermions described by a quadratic Hamiltonian~\cite{jordan1928,sachdev_2011}. The transformation that allows us to do this is a non-local mapping known as the Jordan-Wigner (JW) transformation~\cite{jordan1928}. By working in momentum space, one can see that the Hamiltonian creates pairs of excitations with opposite momenta, which is known as a P-wave superconductor~\cite{Kitaev2001}. This effectively allows us to decompose the dynamics in terms of independent two-level systems in the particle hole basis~\cite{sachdev_2011}.

\subsection{Bogoliubov the Gennes Hamiltonian and pseudo-spin representation }
 
After applying the JW transformation and the discrete Fourier transformation
$\hat{f}_j=\frac{e^{-\mathrm{i}\frac{\pi}{4}}}{\sqrt{N}}\sum_k \hat{F}_{k} e^{\mathrm{i}k j}$ to the Ising model in Eq.~\eqref{eq:IsingHamitonian}, we obtain a fermionic Hamiltonian~\cite{Dziarmaga2005,sachdev_2011},
\begin{align}
          \label{eq:IsingHamitonianFermion}
\hat{H}(t)
=\sum_{k\geq 0} \boldsymbol{\hat{\Psi}}_k^{\dagger}\boldsymbol{H}_k\boldsymbol{\hat{\Psi}}_k
\ ,
\end{align}
where  $\boldsymbol{\hat{\Psi}}_k^{\dagger}=(\hat{F}^{\dagger}_{k},\hat{F}_{-k})$. In appendix~\ref{AppendixA} we provide a detailed derivation of Eq.~\eqref{eq:IsingHamitonianFermion}. The matrix representation 
\begin{align}
          \label{eq:BdG}
\boldsymbol{H}_k=2\hbar[g(t)-J(t)\cos k]\sigma_z+2\hbar J(t)\sin k\ \sigma_x
\end{align}
of the fermionic quadratic form is known as the Bogoliubov de Gennes Hamiltonian and describes a one-dimensional P-wave superconductor~\cite{Kitaev2001}. Here $\sigma_x,\sigma_y$, and $\sigma_z$ are Pauli matrices in the particle-hole basis.
Importantly, as the Hamiltonian is quadratic the Heisenberg equations of motion are linear and can be written in terms of the entries of the Bogoliubov de Gennes Hamiltonian as follows

\begin{align}
          \label{eq:BogoliuvobEq}
         \mathrm{i} \frac{d }{d t}
\begin{bmatrix}
   \hat{F}_{k} \\
    \hat{F}^{\dagger}_{-k}  \\
    \end{bmatrix}
= 
\begin{bmatrix}
     2(g(t)-J(t)\cos k) & 2J(t)\sin k   \\
    2J(t)\sin k  & - 2(g(t)-J(t)\cos k)  \\
    \end{bmatrix}
    .\begin{bmatrix}
    \hat{F}_{k}  \\
   \hat{F}^{\dagger}_{-k}  \\
    \end{bmatrix}
\ ,
\end{align}
which has a general solution $\boldsymbol{\hat{\Psi}}_k(t)=\boldsymbol{U}_{k}(t)\cdot\boldsymbol{\hat{\Psi}}_k(0)$, where
\begin{align}
          \label{eq:FermionunitaryDyn}
\boldsymbol{U}_{k}(t)
= 
\begin{bmatrix}
    \mathcal{U}_k(t) &  \mathcal{V}_k^*(t)   \\
   \mathcal{V}_k(t)  & \mathcal{U}_k^{*}(t)  \\
    \end{bmatrix}
\end{align}
is a propagator for the operators in the Heisenberg picture~\cite{Dziarmaga2005}. In appendix~\ref{AppendixB} we provide a detailed explanation of the relation between the evolution of the fermionic operators in the Heisenberg picture and the explicit mapping to spin states in the Sch\"{o}dinger picture.

\subsection{QSP for fermionic operators in the Heisenberg picture} 
At the formal level, now we can use the propagator of the fermionic operators in Eq.~\eqref{eq:FermionunitaryDyn} to do QSP in the Heisenberg picture. The advantage that we have of working in this framework is that we effectively reduce the problem of the infinite-dimensional Onsager algebra to an effective su$(2)$ algebra in the Heisenberg picture. 
In fact, from the general QSP protocol defined in Eq.~\eqref{eq:IsingQSP}, we can construct a QSP protocol in the Heisenberg picture by using the Bologiubov de Gennes Hamiltonian in Eq.~\eqref{eq:BdG} as follows
\begin{align}     
 \label{eq:BogIsingQSP}
\boldsymbol{U}_{k,\vec{\phi}}(\theta)&=e^{-2\mathrm{i}\phi_{0}\sigma_z}
\prod^d_{r=1}e^{2\mathrm{i} \theta (\sigma_z \cos k-\sigma_x \sin k)}
e^{-2\mathrm{i}\phi_{r}\sigma_z}
\ .
\end{align}
This iterative gate sequence resembles the conventional QSP protocol. However, in order to use the conventional QSP methods to design and analyze the action of the gate sequence in the fermionic mode space, we need to identify the signal and processing unitaries~\cite{Low2017} associated with the proposed gate sequence.

It is worth mentioning that the Krammers-Wannier duality also has a representation in terms of Bologiubov de Gennes Hamiltonian in Eq.~\eqref{eq:BdG}. We can show that the dual QSP in Eq.~\eqref{eq:IsingDualQSP} is obtained by exchanging the order of the operations and roles of the parameters $\phi_{r}$ and $\theta$ in Eq.~\eqref{eq:BogIsingQSP}, as follows
\begin{align}
          \label{eq:DualBogIsingQSP}
\boldsymbol{U}^{\text{D}}_{k,\vec{\phi}}(\theta)&=e^{2\mathrm{i} \phi_{0} (\sigma_z \cos k-\sigma_x \sin k)}\prod^d_{r=1}e^{-2\mathrm{i}\theta\sigma_z}e^{2\mathrm{i} \phi_{r} (\sigma_z \cos k-\sigma_x \sin k)}
\ .
\end{align}
The Krammers-Wannier duality becomes extremely simple in the Heisenberg picture when we use the particle-hole basis. In fact, the QSP protocols in Eqs.~\eqref{eq:DualBogIsingQSP}~and~\eqref{eq:BogIsingQSP} are related by the combined action of a rotation and complex conjugation, as follows
\begin{align}
          \label{eq:KrammersWannierBogIsingQSP}
\boldsymbol{U}^{\text{D}}_{k,\vec{\phi}}(\theta)=
e^{-\mathrm{i}\frac{k}{2}\sigma_y} \boldsymbol{U}^{*}_{-k,\vec{\phi}}(\theta)e^{\mathrm{i}\frac{k}{2}\sigma_y}
\ .
\end{align}
This relation resembles Eq.~\eqref{eq:OperationalDuality} for the quantum circuits shown in Fig.~\ref{Fig1}.

\subsection{Expressivity of QSP in the Heisenberg picture} 
The main difference between the gate sequence in Eq. (\ref{eq:BogIsingQSP}) and the usual qubitization/QSP setup is that in the proposed scheme, the rotation axes of the two single-qubit rotations in each iteration are not orthogonal to one another. Moreover, the angle between the two rotation axes depends on the momentum $k$ of the fermionic mode. Consequently, the identification of the signal and processing unitaries is not immediate. However, this problem can be resolved by noticing the following identity for the $k$ dependent generator SU$(2)$ rotations 
\begin{align}
e^{2\mathrm{i} \theta (\sigma_z \cos k-\sigma_x \sin k)} = e^{i\frac{\pi}{4} \sigma_z }e^{-i\frac{k}{2}\sigma_x} e^{i2\theta \sigma_z} e^{i\frac{k}{2} \sigma_x} e^{-i\frac{\pi}{4} \sigma_z }
\ .
\end{align}
From this identity we obtain the QSP sequence
\begin{align}
\boldsymbol{U}_{k,\vec{\phi}}(\theta)=e^{i(\pi/4-2\phi_{0}) \sigma_z }\left(\prod_{r=1}^{d}e^{-i\frac{k}{2} \sigma_x} e^{i2\theta \sigma_z} e^{i\frac{k}{2} \sigma_x}e^{-2\mathrm{i}\phi_{r}\sigma_z}\right)e^{-i\frac{\pi}{4} \sigma_z }
\label{eq:QSVT}
 \ .
\end{align}
The sequence in parentheses is identical to to the QSVT scheme in Ref.~\cite{Gilyen2019}, except that the phase sequence is constrained by $\theta$.  
The data processed with QSP are encoded in the projected unitary
\begin{align}
\ket{0}_k\bra{0}_k e^{i\frac{k}{2} \sigma_x} \ket{0}_k\bra{0}_k= \cos{(k/2)}\ket{0}_k\bra{0}_k.
\end{align}
The achievable set of polynomial functions of $\cos{(k)}$ using the constrained phase sequence is smaller than that of standard QSVT. First, it is clear that only even parity functions of the signal can be implemented. Otherwise, the constraints seem to be not very strong. 

We first show that when $\theta = \pi/4$, the evolution of the fermionic creation and annihilation operators for each momentum sector can be simplified.   To obtain the desired simplification, first consider taking $\sigma_z$ as the generator of the processing unitary. Then the the block-encoded signal is $\cos{(2\theta)} + \mathrm{i} \cos{(k)}\sin{(2\theta)}$ because
\begin{align}
 e^{2\mathrm{i} \theta (\sigma_z \cos k-\sigma_x \sin k)}= \cos (2\theta)\hat{1}+\mathrm{i}(\sigma_z \cos k-\sigma_x \sin k)\sin (2\theta) 
 \ .
\label{eq:Xevolve}
\end{align}
Crucially, the block encoded signal is $\cos{(k)}$ when $\theta = \pi/4$. Physically, this value allows to create maximally-entangled states in arrays of qubits via the Ising interaction~\cite{Briegel2001}. In terms of experimental implementations, this value of $\theta$ is within reach in currently available arrays of superconducting qubits~\cite{mi2022time}.

Next, we discuss in more detail the special case mentioned above.  By inspecting Eq.~\eqref{eq:QSVT}, we see that if we set $\theta=\pi/4$ 
in Eq.~\eqref{eq:QSVT} we obtain the QSP sequence in the canonical form
\begin{align}
\boldsymbol{V}_{k,\vec{\Phi}}&=
e^{i(\pi/4-2\phi_{0}) \sigma_z }\left(\prod_{r=1}^{d}e^{-ik \sigma_x}  e^{\mathrm{i}(\pi/2-2\phi_{r})\sigma_z}\right)e^{-i\frac{\pi}{4} \sigma_z }
\nonumber \\ &=
e^{i\Phi_{0} \sigma_z }\prod_{r=1}^{d}e^{-ik \sigma_x} e^{\mathrm{i}\Phi_{r}\sigma_z}
\label{eq:QSVT_w_fixed_angle}
\ ,
\end{align}
where the signal operator is a rotation along $x$-axis with an angle proportional to the quasimomentum $k$. The signal processing can be accomplished through a sequence of rotations along the $z$-axis by new angles defined as
\begin{align}
\label{eq:ShiftedQSP} 
        \vec{\Phi}&=(\Phi_0,\Phi_1,\Phi_2,\dots,\Phi_{d-1},\Phi_d)
\ ,
\end{align}
where this sequence is obtained by defining the endpoints phases $\Phi_0=\pi/4-2\phi_0$ and $\Phi_d=\pi/4-2\phi_d$ and $\Phi_r=\pi/2-2\phi_r$ for $r=1,\dots,d-1$, where $\phi_r$ are the phases of the original sequence  $\vec{\phi}=(\phi_0,\phi_1,\dots\phi_d)$.

For convenience, from now on in our paper we use the notation $\boldsymbol{V}_{k,\vec{\Phi}}=\boldsymbol{U}_{k,\vec{\Phi}}(\pi/4)$ to distinguish this special unitary. We will also use $\hat{V}^{\text{O}}_{\vec{\phi}}=\hat{U}_{\vec{\phi}}(\pi/4)$ to denote the corresponding QSP sequence in terms of the Onsager algebra. Later on, we will provide examples to highlight the importance of $\boldsymbol{V}_{k,\vec{\Phi}}$ for applications.

As the signal and signal processing operator are rotations along orthogonal axis, we can use standard techniques and exploit Eq.~\eqref{eq:SignalProcessingOperatorQubit} to obtain QSP sequence for $\vec{\Phi}$
\begin{align}
          \label{eq:QSPExpressability}
\boldsymbol{V}_{k,\vec{\Phi}}=\begin{bmatrix}
    P(x_k) &  \mathrm{i} Q(x_k) \sqrt{1-x_k^2}  \\
     \mathrm{i}Q^*(x_k) \sqrt{1-x_k^2}  & P^*(x_k)  \\
    \end{bmatrix}
\ .
\end{align}
From this it follows that any (bounded, definite parity) polynomial of $x_k=\cos{(k)}$ can be implemented. In turn, for $\theta = \pi/4$, the QSP protocol achieves an optimal expressivity for all the values of $k$ because the axis for the signal and signal processing rotations are orthogonal. Moreover, as the QSP sequence Eq.~\eqref{eq:QSVT_w_fixed_angle} and its dual in Eq.~\eqref{eq:DualBogIsingQSP} are related via Eq.~\eqref{eq:KrammersWannierBogIsingQSP}, the dual QSP sequence also exhibits a high expressivity for $\theta=\pi/4$. This,  follows from Eq.~\eqref{eq:KrammersWannierBogIsingQSP} because the the Y-rotations can be further decomposed into Z-conjugated X rotations according to $k$ and this means that the dual protocols have the same form as the original protocols, with the addition of one additional iterate (signal oracle). This asymmetry is due to the fact that the general QSP protocol has $d$ signal operators and $d + 1$ controllable phases.
\begin{remark}
QSP is mainly a statement about the mathematical form of a product of parameterized SU$(2)$ operations. Usually we denote the signal by $\theta$, and consider it an unknown~\cite{Low2017,Martyn2021}, but whenever an unknown appears and parameterizes such a product, it can be treated in place of $\theta$. In the QSP sequence $\boldsymbol{U}_{k,\vec{\phi}}(\theta)$ of Eq.~\eqref{eq:QSVT}, a new variable (the momentum $k$) appears given our problem statement. As we have multiple choices for the signal, in some situations it makes sense to tune the (known, and thus controllable) $\theta$ dependence, effectively removing it by setting $\theta=\pi/4$, and leaving the momentum to be processed within each  subspace labelled by $k$. In the general case, one can still use Eq.~\eqref{eq:QSVT} when $\theta$ is unknown, but one has to determine the expressivity a two-variable QSP sequence with not orthogonal axis. In appendix~\ref{AppendixC} we discuss a modified QSP sequence for arbitrary $\theta$ and $k$ in such a way that the signal and signal processing operations are rotations along orthogonal axis. In contrast to the usual QSP, the axis of the signal operator is defined by $k$ and $\theta$ in a nonlinear fashion. This is of course an interesting problem by itself, but it is beyond the scope of our current work.
\end{remark}
\section{Applications and examples of QSP with the Onsager algebra\label{SecVI}} 
At this stage it is important to consider some particular examples to see how QSP works in the Heisenberg picture by using the QSP sequence $\boldsymbol{V}_{k,\vec{\Phi}}$ of Eq.~\eqref{eq:QSVT_w_fixed_angle} in momentum space with angles $\vec{\Phi}$. As we discussed above, in some cases, it is useful to fix $\theta=\pi/4$ to treat the momentum $k$ as the signal to be processed. This particular value of $\theta$ is extremely important for applications as it allows the maximum expressivity for QSP sequences in momentum space.
We will start with an example where we discuss the trivial QSP sequence. In the second example, we discuss QSP sequences for the Onsager algebra and the relation to space-time dual quantum circuits, which are relevant in quantum information processing and in the study of quantum signatures of manybody chaos~\cite{Akila2016,Piroli2020,Bertini2018,Lu2021,Fisher2023}. The next two examples are related to the use of our scheme for quantum simulation of Hamiltonians. The last example reframes a well-known protocol in NMR to synthesize a BB1 sequence~\cite{Wimperis1994} for the Onsager algebra~\cite{Onsager1944,Gritsev2017}.

\subsection{Trivial QSP sequence in momentum space}
The simplest example of a QSP sequence can be obtained by considering $\vec{\Phi}=(0,0,0)$ in Eq.~\eqref{eq:QSVT_w_fixed_angle}. This gives us the trivial QSP sequence in momentum space
\begin{align}
          \label{eq:FirstExampleBogIsingQSP}
\boldsymbol{V}_{k,\vec{\Phi}}=e^{-i2 k \sigma_x} 
\ .
\end{align}
From this, we obtain the associated polynomial transformation of the input $P(x_k) =2x^2_k-1$. Similarly, for $\vec{\Phi}=(0,0,0,0)$ we obtain $P(x_k) =4x^3_k-3x_k$. For a trivial protocol with length $d$, one can show that the resulting polynomial transformation is given by the Chebyshev polynomials of the first kind $P(x_k) =T_d(x_k)$ as in Ref.~\cite{Martyn2021}. The purpose of this example is to show the versatility of Eq.~\eqref{eq:QSVT_w_fixed_angle}. As this has the canonical form of the QSP known in the literature, we can use it to analyze QSP sequences with rotations $\vec{\Phi}$ in momentum space. Then, we can translate those back into angles $\vec{\phi}$ defining the corresponding QSP sequence for the Onsager algebra. For example, in the case of $\vec{\Phi}=(0,0,0,0)$, the original angles are given by 
\begin{align}
\label{eq:TrivialSeq} 
     \vec{\phi}&=(\pi/8,\pi/4,\pi/4,\pi/8)
        \ ,
\end{align}
and define the QSP sequence $\hat{V}^{\text{O}}_{\vec{\phi}}$ for the Onsager algebra [see Eq.~\eqref{eq:IsingQSP}].

\subsection{Space-time rotation and dual quantum circuits}
Now let us consider a more involved example related to the theory of space-time dual quantum circuits. Motivated by a recent work~\cite{Lu2021}, we consider dual quantum circuit in the absence of disorder. Recently space time duality has attracted much attention, with connections to topics ranging from quantum signatures of manybody chaos~\cite{Akila2016,Fisher2023} to dynamical quantum phase transitions~\cite{hamazaki2021exceptional}. One of the most appealing aspects of this theory is that it allows one to obtain analytical results even when dynamics are ergodic~\cite{Bertini2018}. 

To make the connection between the theory space-time dual quantum circuits and QSP for the Onsager algebra, we can consider the sequence of operations in Eq.~\eqref{eq:IsingQSP} for fixed $\theta=\pi/4$ and  $\vec{\phi}=[0,\frac{\pi}{2}(1-2\epsilon),\frac{\pi}{2}(1-2\epsilon), \ldots,\frac{\pi}{2}(1-2\epsilon)]$, where $\epsilon$ is an error in the rotation angle [see Eq.~\eqref{eq:ShiftedQSP}]. The QSP sequence with $d$ time steps for a lattice with $N$ sites reads
\begin{align}
          \label{eq:IsingPeriodicQSP}
\hat{V}^{\text{O}}_{\vec{\phi}}=\prod^d_{r=1}e^{\mathrm{i} \frac{\pi}{4}\sum^{N}_{j=1}X_j X_{j+1}}
e^{\mathrm{i}\phi_{r}\sum^{N}_{j=1} Z_j}
\end{align}
with $\phi_{r}=\frac{\pi}{2}(1-2\epsilon)$. 

To build a space-time dual QSP, we change the roles of space and time. In other words, the dual QSP sequence corresponds to $N$ iterations in time of a Hamiltonian acting on $d$ sites in space, as follows
\begin{align}
          \label{eq:SpaceTimeIsingDualQSP}
\hat{V}^{\text{DST}}_{\vec{\widetilde{\phi}}}=\prod^{N}_{r=1}e^{\mathrm{i} \widetilde{\phi}_{r}\sum^{d}_{j=1} \tilde{Z}_j}
e^{\mathrm{i}\widetilde{\theta}\sum^{d}_{j=1}\tilde{X}_j \tilde{X}_{j+1}}
\ ,
\end{align}
where $ \widetilde{\phi}_{r}=-\pi/4$ and $\widetilde{\theta}=-\pi/4+\mathrm{i}/2\log\{\tan[\pi/2(1-2\epsilon)]\}$~\cite{Lu2021}. We note that this has the same form as dual Onsager QSP sequence in Eq.~\eqref{eq:IsingDualQSP}. The main difference is that the Krammers-Wannier duality exchanges the roles of signal and signal processing sequence, while keeping the evolution unitary~\cite{2017Motrunich}. Under the space-time duality, however, the QSP sequence is not unitary. In terms of the parameter $\epsilon$, there is a special value $\epsilon=1/4$ for which the dual quantum circuit is unitary and $\widetilde{\theta}=-\pi/4$.

\begin{figure}
	\includegraphics[width=0.45
\textwidth]{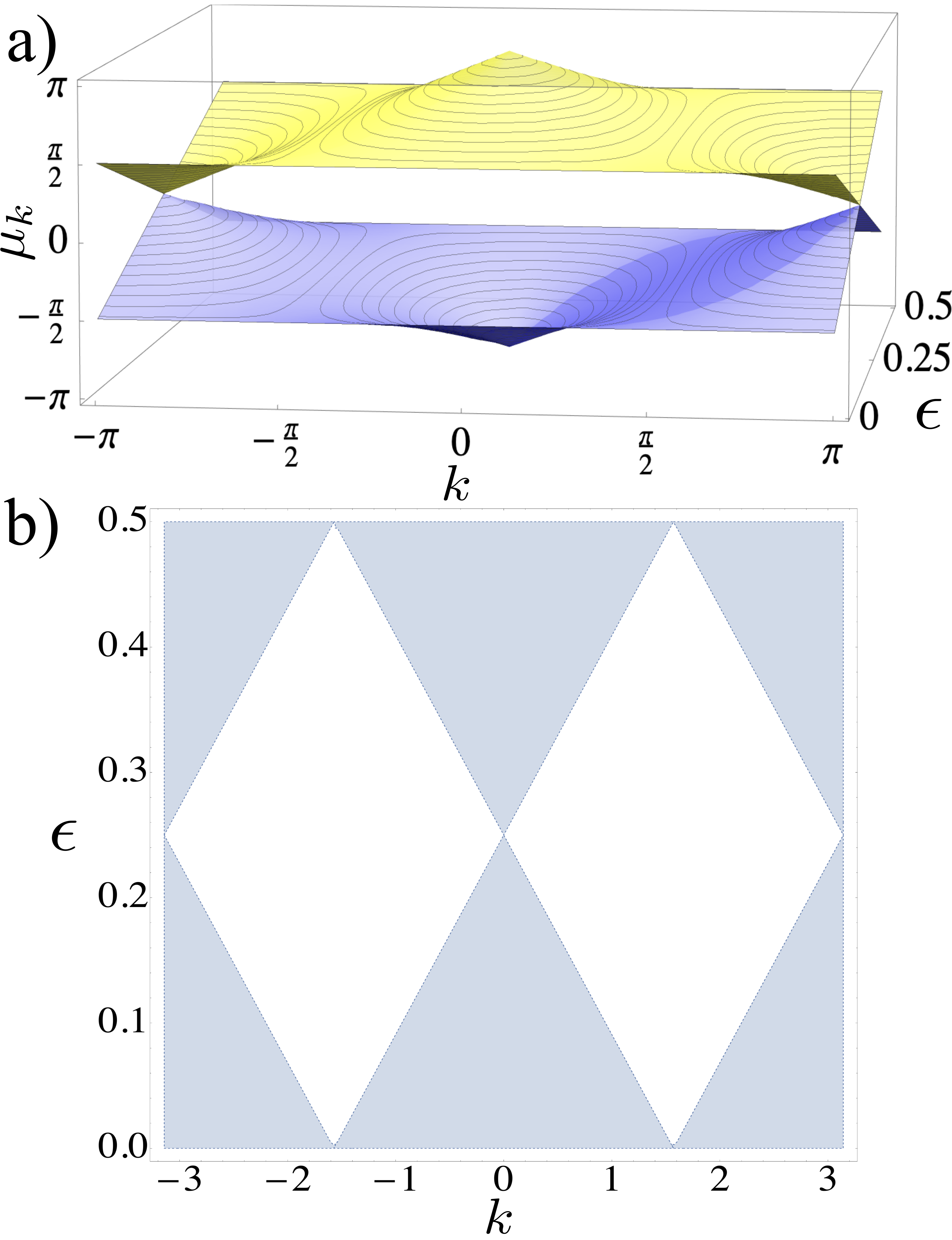}
	\caption{Spectral properties of a QSP sequence and its space time dual. a) Depicts the Floquet exponent $\mu_k$ of the iterator as a function of the error and the quasimomentum $k$. There are both $0$- and $\pi$-gaps and the Floquet exponents are independent on the error for $k=\pi/2$, as we predicted using QSP methods. The critical point at $\epsilon=1/4$ are ensured by space-time duality (see main text). b) Depicts the phase diagram determining the features of the spectrum of the space time dual QSP sequence. For parameters within the white region the eigenvalues satisfy the condition $|\lambda^{\text{DST}}_k|=1$ and the evolution is unitary. For the self-dual point  $\epsilon=1/4$ of the space time dual quantum circuit there is a singularity at momenta $k=0$ and $|k|=\pi$ in correspondence with the gapless excitation spectrum in shown in panel~a).}
	\label{Fig2}
\end{figure}

Next, let us explore some properties of the QSP sequence in Eq.~\eqref{eq:IsingPeriodicQSP} by working in quasimomentum space
\begin{align} 
\label{eq:BogIsingQSPR}
\boldsymbol{V}_{k,\vec{\Phi}}&=\prod^d_{r=1}e^{\mathrm{i} \frac{\pi}{2} (\sigma_z \cos k-\sigma_x \sin k)}
e^{-2\mathrm{i}\pi(1-2\epsilon)\sigma_z}
\ ,
\end{align}
As the QSP protocols involve constant phases, at each time step the evolution is given as a product of two unitaries. Thus by using Floquet theory, we can extract most relevant information from the evolution operator in one period of the sequence, defining the Floquet operator
\begin{align} 
\label{eq:FloquetBog}
\mathcal{F}_k=e^{\mathrm{i} \frac{\pi}{2} (\sigma_z \cos k-\sigma_x \sin k)}
e^{-\mathrm{i}\pi(1-2\epsilon)\sigma_z}
\ .
\end{align}
The eigenvalues of the Floquet operator are $\lambda_k=\exp(-\mathrm{i}\mu_k)$ and $\mu_k$ are the Floquet exponents. 
 For example when $k=0,\pi$, the Floquet exponents are $\mu_{0}=\pi-2\pi|\epsilon-1/4|$ and $\mu_{\pi}=2\pi|\epsilon-1/4|$. 
When $\epsilon_c=1/4$, there is a $\pi$-energy gap for $k=0$ and a zero energy gap for the mode $k=\pm\pi$ indicating a quantum critical point at $\epsilon_c$ that is the self-dual point under space-time duality~\cite{Lu2021}. For quasimomentum $k=\pi/2$ the Floquet exponent is independent of the error and is given by $\mu_{\pi/2}=\pi/2$. In Appendix~\Ref{AppendixD} we discuss the QSP sequence in for $k=\pi/2$. Figure \ref{Fig2}~a) shows the Floquet exponents $\mu_{k}$ as a function of the quasimomentum and the error. From this we can see the $0$- and $\pi$-gaps indicating the sefl-dual point.

To obtain more information about the space-time dual QSP sequence in Eq.~\eqref{eq:SpaceTimeIsingDualQSP}, we consider the momentum representation
\begin{align}
\boldsymbol{V}^{\text{DST}}_{k,\vec{\widetilde{\phi}}}=\prod^N_{r=1}e^{\frac{\mathrm{i}\pi}{2}\sigma_z}e^{2\mathrm{i} \widetilde{\theta} (\sigma_z \cos k-\sigma_x \sin k)}
\label{eq:QSVTDual}
 \ .
\end{align}
Similarly to the QSP sequence in Eq.~\eqref{eq:ThetapiNew} discussed above, due to the periodicity, it is enough to study spectral properties of the non-unitary version of the Floquet operator
\begin{align} 
\label{eq:FloquetBog}
\mathcal{F}^{\text{DST}}_k=e^{\frac{\mathrm{i}\pi}{2}\sigma_z}e^{2\mathrm{i} \widetilde{\theta} (\sigma_z \cos k-\sigma_x \sin k)}
\ .
\end{align}
In contrast to its unitary version, the eigenvalues $\lambda^{\text{DST}}_k$ of the Floquet operator $\mathcal{F}^{\text{DST}}_k$ are not restricted to lie along the unit circle. In fact, depending on the momentum $k$ and the error $\epsilon$, they may satisfy $|\lambda^{\text{DST}}_k|<1$ or $|\lambda^{\text{DST}}_k|>1$. Figure~\ref{Fig3} depicts a region plot in the $k-\epsilon$ parameter space  where the white region is determined by the condition of unitarity $|\lambda^{\text{DST}}_k|=1$. Interestingly, and as we discussed below, the momentum $|k|=\pi/2$ lies in the white region for all values of the error and there is correspondence between the $0-$ and $\pi-$ gaps in Fig.~\ref{Fig2}~a) and the behavior of the line $\epsilon=1/4$. In fact, in the shaded region, each eigenvalue satisfying $|\lambda^{\text{DST}}_k|<1$ has an exact partner such as $|\lambda^{\text{DST}}_k|>1$. That being said, some modes are amplified~\cite{Basu2022} and others are suppressed for parameters within the shaded region in Fig.~\ref{Fig2}~b). This spectral properties have important consequences. For example, due to long-lived quasiparticle pairs with purely real energy, the dual quantum circuit reaches a steady state with volume-law
entanglement~\cite{Lu2021}.

\subsection{Design of pulse sequences to simulate the response under a target spin Hamiltonian} 
In the previous sections, we have been focusing on describing the general formalism for QSP in terms of the spin representation of the Onsager algebra and in the Heisenberg picture. In this subsection, we will provide an example of a possible application of QSP to simulate a Hamiltonian by designing a pulse sequence. With this aim, let us consider the a simple target Hamiltonian of the form
\begin{align}
          \label{eq:ClusterHamitonian}
\hat{H}_{\text{Target}}(t)&=-\hbar g_0\sum^{N}_{j=1} Z_j-\hbar J_x\sum^{N}_{j=1}X_j Z_{j+1}Z_{j+2}Z_{j+3}X_{j+4}
\nonumber \\
&
-\hbar J_y\sum^{N}_{j=1}Y_j Z_{j+1}Z_{j+2}Z_{j+3}Y_{j+4}
\ .
\end{align}
It is convenient to introduce the notation $J_x=J(1+\gamma)/2$ and $J_y=J(1-\gamma)/2$, where $\gamma$ is a dimensionless parameter characterizing the anisotropy of the interaction.
\begin{figure}
	\includegraphics[width=0.45
\textwidth]{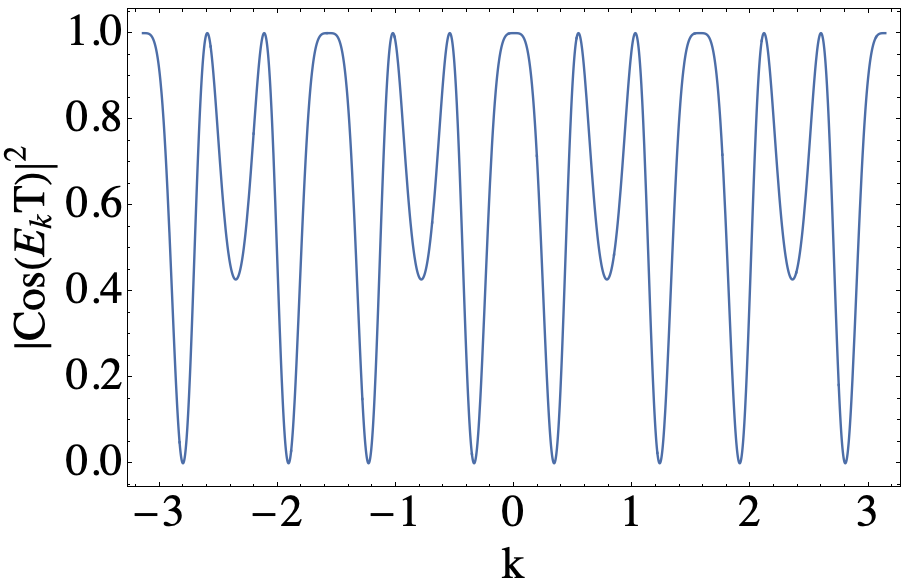}
	\caption{Probability $|\bra{+}_k\boldsymbol{U}_{k}\ket{+}_k|^2=\cos^2 (E_kT)$  corresponding to the cluster Hamiltonian Eq.~\eqref{eq:ClusterHamitonian} in momentum space. We set parameters $g=J$ and $\gamma=0$.}
	\label{Fig3}
\end{figure}
Certainly, it is a nontrivial task to find a sequence of rotations $\vec{\phi}$ in such a way that the resulting unitary $\hat{U}_{\vec{\phi}}(\theta)$ from the QSP sequence in the spin representation of Eq.~\eqref{eq:IsingQSP} is close to our desired target Hamiltonian for arbitrary $\theta$. As the algebra is infinite dimensional in the limit $N\rightarrow\infty$, the number of commutators required makes the procedure impractical. However, as we will show below, one can obtain an enormous simplification of the problem in the Heisenberg picture in the fermionic representation when we set $\theta=\pi/4$ and work with the QSP sequences in Eqs.~\eqref{eq:QSVT_w_fixed_angle}~and~\eqref{eq:QSPExpressability}.

By applying the Jordan-Wigner transformation and the discrete Fourier transformation of the fermionic operators as we did in the case of the Ising chain, we can obtain the Bogoliubov de Gennes Hamiltonian 
\begin{align}
          \label{eq:ClusterBdG}
\boldsymbol{H}^{\text{Target}}_k
     =2\hbar[g_0-J\cos 4k]\sigma_z+2\hbar J\gamma\sin 4k\sigma_x
\ .
\end{align}
corresponding to Eq.~\eqref{eq:ClusterHamitonian}. 
We can rewrite this in the form $\boldsymbol{H}^{\text{Target}}_k=\hbar \Omega_k \boldsymbol{n}_k\cdot \boldsymbol{\sigma} $, where  $\boldsymbol{\sigma}=(\sigma_x,\sigma_y,\sigma_z)$ and
\begin{align}
      \label{eq:ClusterDispersion}
\Omega_k&=2\sqrt{[g_0-J\cos 4k]^2+(J\gamma)^2\sin^2 4k}
\nonumber\\
\boldsymbol{n}_k&=\frac{2[g_0-J\cos 4k]}{\Omega_k}\sigma_z+\frac{2J\gamma\sin 4k}{\Omega_k}\sigma_x
\ .
\end{align}
This defines $n_x(k)=2J\gamma\sin 4k/\Omega_k$ and $n_z(k)=2[g_0-J\cos 4k]/\Omega_k$.
After an evolution time $T$, the quantum evolution under $\boldsymbol{H}^{\text{Target}}_k$ is given by the unitary operator
\begin{align}
    \label{eq:EvolutionClusterBdG}
\boldsymbol{U}_{k}=\begin{bmatrix}
   \cos (\Omega_kT)-\mathrm{i}n_z(k)\sin (\Omega_kT) & -\mathrm{i}n_x(k)\sin (\Omega_kT)  \\
     -\mathrm{i}n_x(k)\sin (\Omega_kT)  & \cos (\Omega_kT)+\mathrm{i}n_z(k)\sin (\Omega_kT)  \\
         \end{bmatrix}
\ .
\end{align}
From this we can see that the matrix elements are functions that could be approximated using QSP in the Heisenberg picture. That is, there is a sequence $\vec{\Phi}$ that acts as a polynomial transformation of the input
\begin{align}
    \label{eq:QSimApprox}
\bra{0}_k\boldsymbol{U}_{k}\ket{0}_k&=\cos (\Omega_kT)-\mathrm{i}n_x(k)\sin (\Omega_kT)
\nonumber \\
&\approx\bra{0}_k\boldsymbol{V}_{k,\vec{\Phi}}\ket{0}_k \ ,
\end{align}
where $\boldsymbol{V}_{k,\vec{\Phi}}$ was defined in Eqs.~\eqref{eq:QSVT_w_fixed_angle}~and~\eqref{eq:QSPExpressability}. In the previous discussion we faced a restriction when the signal $x_k=\cos(k)=\pm1$, or equivalently, when $k=0,\pi$. In this case the signal is proportional to the identity and the QSP sequence turns out to be a single Z-rotation. Keeping this in mind, in terms of the numerical implementation we can accurately approximate the function
\begin{align}
    \label{eq:QSimApprox}
\bra{+}_k\boldsymbol{U}_{k}\ket{+}_k=\cos (\Omega_kT)\approx\bra{+}_k\boldsymbol{V}_{k,\vec{\Phi}}\ket{+}_k \ ,
\end{align}
where $\ket{+}_k=(\ket{0}_k+\ket{1}_k)/\sqrt{2}$. Figure~\ref{Fig3} shows the behavior of this response function for different values of $k$. The expressivity of the QSP sequence in the standard form of Eq.~\eqref{eq:QSPExpressability} has been widely investigated. Therefore, there are efficient ways to obtain a sequence of phases $\vec{\Phi}$ that gives us a good polynomial approximation to a desired function. In turn, this sequence of operations can be used to design a QSP sequence $\hat{V}^{\text{O}}_{\vec{\phi}}$ in terms of the original Pauli operators to simulate the action of the evolution operator Eq~\eqref{eq:EvolutionClusterBdG} that is generated by the Hamiltonian Eq.~\eqref{eq:ClusterBdG}. 


\subsection{Reverse engineering of spin Hamiltonians from response functions in momentum space} 
In this subsection let us present another example example based on the idea of reverse engineering spin Hamiltonians from a given polynomial transformation in momentum space. For simplicity, we consider a phase sequence that has a simple limiting behavior in momentum space and then show that there is a pre-image spin Hamiltonian in real space which would induce this evolution. 

As a starting point to construct our example, we assume a simple form for the unitary evolution
\begin{align}
    \label{eq:ReverseEngineer}
\boldsymbol{U}_{k}=e^{-\mathrm{i}  \Omega_kT \sigma_x }
=\begin{bmatrix}
   \cos (\Omega_kT) & -\mathrm{i}\sin (\Omega_kT)  \\
     -\mathrm{i}\sin (\Omega_kT)  & \cos (\Omega_kT) \\
         \end{bmatrix}
\ .
\end{align}
Clearly, the response function associated to this evolution is given by $\bra{0}_k\boldsymbol{U}_{k}\ket{0}_k=\cos (\Omega_kT)$. We can think of defining a ``reversed engineered" Hamiltonian
$\boldsymbol{H}^{\text{RE}}_{k}=\hbar \Omega_k \sigma_x$. For concreteness, we will focus here on an example provided in the appendix D of Ref.~\cite{Martyn2021} of a phase sequence as a polynomial approximation for phase estimation function.
\begin{align}
    \label{eq:RectResponse}
 \cos (\Omega_kT)=2\Pi(3x_k/2)-1
 \ .
\end{align}
where $\Pi(z)$ denotes the box distribution (also known as the Heaviside Pi function). It follows that the angular frequency dispersion $\Omega_k=\pi/T-\pi\Pi(3x_k/2)/T$. We now employ Fourier analysis to obtain the expression
\begin{align}
    \label{eq:FourierTrick}
 \Pi\left(\frac{3x_k}{2}\right)=\frac{3}{4\pi}\int^{\infty}_{-\infty}\frac{\sin(3\omega/4)}{3\omega /4}e^{\mathrm{i}\omega x_k}d\omega
=\frac{3}{4\pi}\sum_{n=-\infty}^{\infty}\mathrm{i}^n e^{\mathrm{i}n k}G_n
\ ,
\end{align}
where $G_{n}=\int^{\infty}_{-\infty}\sin(3\omega /4)/(3\omega /4)\mathcal{J}_{n}(\omega)d\omega$ with $\mathcal{J}_n(\omega)$ being a Bessel function of the first kind~\cite{gradshteyn2014table}. From these relations, we can obtain a closed form for the Hamiltonian
\begin{align}
    \label{eq:ExplicitHamiltonian}
\boldsymbol{H}^{\text{RE}}_{k}=\frac{\hbar}{T}\left[\pi-\frac{3}{2}\left(\sum_{n=0}^{\infty}(-1)^n \cos(2n k)G_{2n}\right)\right]\sigma_x
\ ,
\end{align}
where we have exploited the symmetry $G_{2n}=G_{-2n}$ and the fact that $G_{2n+1}=0$.
With all these elements at hand, we can obtain the fermionic Hamiltonian $\hat{H}^{\text{RE}}=\sum_{k\geq} \boldsymbol{\hat{\Psi}}_k^{\dagger}\boldsymbol{H}^{\text{RE}}_{k}\boldsymbol{\hat{\Psi}}_k$, as follows
\begin{align}
    \label{eq:HamiltonianFermionSpace}
\hat{H}^{\text{RE}}&=-\frac{3\hbar}{2T}\sum_{k\geq0}\sum_{n=0}^{\infty}(-1)^n G_{2n}\left(\cos(2n k)\hat{F}_{-k}\hat{F}_{k}+\text{h.c} \right)
\nonumber\\
&=-\frac{3\hbar \mathrm{i}}{4T}\sum_{j}\sum_{n=0}^{\infty}(-1)^n G_{2n}(\hat{f}_{j-2n}\hat{f}_{j}+\hat{f}_{j+2n}\hat{f}_{j}) +\text{h.c}
\ .
\end{align}
We will not show the derivation here, but the fermionic terms can be re-written in terms of Pauli matrices, giving rise to nonlocal spin Hamiltonians of the form
\begin{align}
    \label{eq:HamiltonianSpinSpace}
\sum_{j}(\mathrm{i}\hat{f}_{j-2n}\hat{f}_{j}+\text{h.c}) =\sum_{j}(X_{j-2n}\hat{M}^z_jX_{j}-Y_{j-2n}\hat{M}^z_jY_{j})
\ .
\end{align}
Here $\hat{M}^z_j=Z_{j-2n+1}\cdots Z_{j-1}$ arises from the Jordan Wigner string connecting the sites $j-2n$ and $j$. We refer to the interested reader to Ref.~\cite{Martyn2021}, that provides the explicit phase sequence required to approximate the phase estimation function.

The example presented above shows there is always some pre-image of a QSP transformation in momentum space in the form of a time-independent spin Hamiltonian in real space that matches the evolution we achieve. However, in general, the pre-image spin Hamiltonian is highly non-local, as we can see from our example. Nevertheless, the QSP sequence in terms of the Onsager algebra $\hat{V}^{\text{O}}_{\vec{\phi}}$ is given as a sequence of single- and two-qubit gates. 

\subsection{BB1 protocol for the quantum Ising chain}
In final subsection our main focus will be to use a paradigmatic composite sequence from the NMR community in the context of our QSP sequence in the momentum space. In turn, our result allows us to define a BB1 protocol for the Onsager algebra applicable to quantum Ising chains.

\begin{figure}
	\includegraphics[width=0.45
\textwidth]{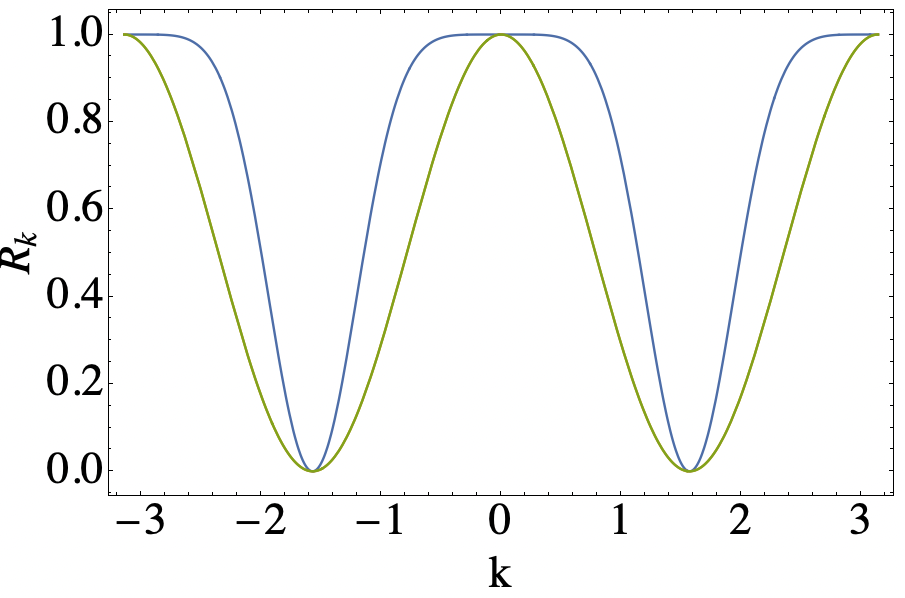}
	\caption{BB1 QSP sequence in momentum space and its effect on the transition probability. The green curve depicts the transition probability $R_k$ in Eq.~\eqref{eq:TrivialSeqProb} without signal processing. The blue curve shows the transition probability $R_k^{\text{BB1}}$ in Eq.~\eqref{eq:BB1SeqProb} after applying the BB1 sequence.}
	\label{Fig4}
\end{figure}

To start, let us consider the QSP sequence $\boldsymbol{V}_{k,\vec{\Phi}}$ in Eq.~\eqref{eq:QSVT_w_fixed_angle} for a fixed angle $\theta=\pi/4$. Notably, if we forget the physical meaning of the quasimomentum $k$, we can interpret it as a signal and the QSP sequence has the same structure as the canonical form of QSP sequence for SU$(2)$ in Eq.~\eqref{eq:SignalProcessingOperatorQubit}. Naively, we can use known QSP sequences for su$(2)$ in the literature to ``synthesize" new QSP sequences for the Onsager algebra.

For concreteness, let us consider a paradigmatic composite pulse sequence in NMR known as the ``BB1" sequence~\cite{Wimperis1994,Martyn2021}. In the context of our QSP sequence in momentum space, we can do some signal processing of the quasimomentum $k$, by considering a sequence of rotations 
\begin{align}
\label{eq:BB1} 
        \vec{\Phi}_{\text{BB1}}=\left(\pi/2,-\chi,2\chi,0,-2\chi,\chi\right)
\ ,
\end{align}
where $\chi=1/2\cos^{-1}(-1/4)$. This has exactly the same form as the BB1 ``composite-pulse" sequence used in NMR. From Eq.~\eqref{eq:ShiftedQSP} and \eqref{eq:BB1} we can retrieve the original phases
\begin{align}
\label{eq:OriginalPhaseBB1} 
        \vec{\phi}_{\text{BB1}}=\left(-\frac{\pi}{8},\frac{\chi}{2}+\frac{\pi}{4},-\chi+\frac{\pi}{4},\frac{\pi}{4},\chi+\frac{\pi}{4},-\frac{\chi}{2}+\frac{\pi}{8}\right)
\end{align}
which allows us to define the BB1 sequence for the Onsager algebra $\hat{V}^{\text{O}}_{ \vec{\phi}_{\text{BB1}}}=\hat{U}_{ \vec{\phi}_{\text{BB1}}}(\pi/4)$ in Eq.~\eqref{eq:IsingQSP}.
In momentum space, the signal to be processed is the momentum $k$ and we can define a QSP sequence $\boldsymbol{V}_{k,\vec{\Phi}_{\text{BB1}}}$ as in Eq.~\eqref{eq:QSVT_w_fixed_angle}. To understand the effect of the BB1 sequence, it is illustrative to obtain the probability in the absence of any processing, i.e., for $\vec{\Phi}=(0,0)$ and a given momentum $k$
\begin{align}
\label{eq:TrivialSeqProb} 
        R_k=|\bra{0}_k\boldsymbol{V}_{k,(0,0)}\ket{0}_k|^2=x_k^2
        \ .
\end{align}
Now, if we apply the BB1 sequence, we obtain the modified transition probability
\begin{align}
\label{eq:BB1SeqProb} 
        R_k^{\text{BB1}}&=|\bra{0}_k\boldsymbol{V}_{k,\vec{\Phi}_{\text{BB1}}}\ket{0}_k|^2
        \nonumber \\
        &=\frac{1}{8}x_k^2\left[3x_k^8-15x_k^6+35x_k^4-45x_k^2+30\right]
        \ ,
\end{align}
where $x_k=\cos(k)$. In NMR, the BB1 sequence is known for allowing the two level system to remain unflipped for a wide range of signals. In our case, in a region around $k=0$ and $k=\pi$. This sequence shows a sharp transition for $|k|\approx\pi/3$ and $|k|\approx2\pi/3$. As a consequence, when applying the BB1 sequence, we obtain a high sensitivity to specific values of the momentum $k$. Here it is important to remark that this step function can be made arbitrarily sharp~\cite{Wimperis1994,Martyn2021}. The
main benefit of BB1, besides its historical status, is that the protocol is relatively short, and its achieved polynomial transform is easy to write down.

But what are the consequences of this sensitivity? Well, the QSP sequence keeps both long-wavelength $(k\approx 0)$ and short-wavelength excitations $(k\approx \pi)$ frozen, while it flips excitations with momentum close to $k\approx \pi/2$. That is, if we prepare an initial spin state $\ket{\Psi(0)}=\prod^{\pi}_{k=-\pi}\ket{0}_k=\ket{\uparrow,\uparrow,\dots,\uparrow,\uparrow}$,
we can calculate the probability
\begin{align}
\label{eq:Onsager} 
        R=|\bra{\Psi(0)}\hat{V}_{ \vec{\phi}_{\text{BB1}}}\ket{\Psi(0)}|^2=\prod^{\pi}_{k=-\pi}R_k^{\text{BB1}} =0
             \ .
\end{align}
This turns out to be exactly zero because $P_{\pm\pi/2}^{\text{BB1}}=0$.

\section{Conclusions\label{SecVII}}
In summary, we have investigated QSP protocols for the Onsager algebra, an infinite dimensional Lie algebra that naturally appears in the theory of the Ising model. We have shown that by mapping the Ising model to a system of non-interacting fermions, we can define QSP protocols for the fermionic operators in the Heisenberg picture respecting the su$(2)$ algebra. This naturally allows one to exploit the tools of standard QSP with SU$(2)$ operations. We then applied such sequences to illustrate various examples and applications in diverse fields ranging from space-time dual quantum circuits, quantum engineering of spin Hamiltonians, and composite pulse sequences in spin chains. These examples highlight the wide utility of our approach and how one can translate QSP sequences in momentum space based on su$(2)$ algebra in the Heisenberg picture to well-defined protocols dependent on the Onsager algebra in the Sch\"{o}dinger picture.

There are of course some remaining open questions that are worth exploring. For example, when we start with the Onsager algebra in the Sch\"{o}dinger picture, after a set of transformations, the evolution of the operators in the Heisenberg picture can be entirely described by the standard theory of QSP. For tuned values of system, we reach the optimal expressivity for QSP sequences in momentum space. However, it remains unclear how generalizable this approach is to other systems defined by other algebras and at other tuned points. It would be worthwhile to determine which classes of physical models permit QSP-like control. This could allow one to make statements about the robustness of QSP in the context of condensed matter systems and quantum simulation. For example, it would be interesting to explore QSP sequences in spin chains such as the XXZ model, which cannot be mapped to systems of interacting fermions~\cite{Cabra2004,von1998bosonization}. To deal with this problem, one can use bosonization to map problems of interacting fermions at half-filling to squeezed collective bosonic modes~\cite{Bukov2012}. This will of course require one to use recently developed QSP sequences based on the su$(1,1)$ algebra for continuous variables~\cite{rossi2023}. It would be interesting to explore the use of QSP methods to treat non-integrable models such as high-dimensional version of the TFIM. For example, a two-dimensional lattice can be represented as a family of coupled one-dimensional TFIMs. In certain regimes, our approach for the one-dimensional TFIM can provide a good approximation for a two-dimensional problem. Other possible extension of our work is to investigate QSP sequences in two-band topological insulators and topological superconductors which can be described using a pseudo-spin approach in momentum space~\cite{Qi2011}.


{\it{Acknowledgments.---} }
The authors would like to thank NTT Research
Inc.  for their support in this collaboration. The authors are thankful for fruitful discussions with S. Sugiura. WJM and
VMB acknowledge partial support through the MEXT Quantum Leap Flagship Program (MEXT Q-LEAP) under Grant No. JPMXS0118069605. ZMR was supported in part by the NSF EPiQC program, and ILC was supported in part by the U.S. DoE, Office of Science, National Quantum Information Science Research Centers, and Co- design Center for Quantum Advantage (C2QA) under Contract No. DE-SC0012704

\appendix

\section{Jordan Wigner transformation and P-wave superconductivity\label{AppendixA}}
The Jordan Wigner transformation allows one to represent the Pauli matrices in terms of fermionic operators. This mapping is highly non-local is given by
\begin{align}
          \label{eq:JordanWigner}
 X_j&=(\hat{f}^{\dagger}_j+\hat{f}_j)\prod^{j-1}_{m=1}(1-2\hat{f}^{\dagger}_m\hat{f}_m)
\nonumber \\
Y_j&=-\mathrm{i}(\hat{f}^{\dagger}_j-\hat{f}_j)\prod^{j-1}_{m=1}(1-2\hat{f}^{\dagger}_m\hat{f}_m)
\nonumber \\
Z_j&=1-2\hat{f}^{\dagger}_j\hat{f}_j
\ .
\end{align}
Here the operators $\hat{f}^{\dagger}_j$ and $\hat{f}_j$ are the fermionic creation and annihilation operators in real space satisfying the anticommutation relations $\{\hat{f}_i,\hat{f}^{\dagger}_j\}=\delta_{i,j}$ and $\{\hat{f}_i,\hat{f}_j\}=\{\hat{f}^{\dagger}_i,\hat{f}^{\dagger}_j\}=0$. 

After applying the JW transformation to the Ising model in Eq.~\eqref{eq:IsingHamitonian}, we obtain the fermionic quadratic Hamiltonian
\begin{align}
          \label{eq:IsingHamitonianFermionSI}
\hat{H}(t)&=-\hbar g(t)\sum^{N}_{j=1} (1-2\hat{f}^{\dagger}_j\hat{f}_j)- \hbar J(t)\sum^{N-1}_{j=1}(\hat{f}^{\dagger}_j-\hat{f}_j)(\hat{f}^{\dagger}_{j+1}+\hat{f}_{j+1})
\nonumber \\
&= 2\hbar\sum_{k\geq} (g(t)-J(t)\cos k)(\hat{F}^{\dagger}_k\hat{F}_k-\hat{F}_{-k}\hat{F}^{\dagger}_{-k})
\nonumber \\
&
+ 2\hbar J(t)\sum_{k\geq}\sin k(\hat{F}^{\dagger}_{k}\hat{F}^{\dagger}_{-k}+\hat{F}_{-k}\hat{F}_{k})
\nonumber \\
&=\sum_{k\geq} \boldsymbol{\hat{\Psi}}_k^{\dagger}\boldsymbol{H}_k\boldsymbol{\hat{\Psi}}_k
\ ,
\end{align}
where $\boldsymbol{\hat{\Psi}}_k^{\dagger}=(\hat{F}^{\dagger}_{k},\hat{F}_{-k})$. Here $\hat{F}^{\dagger}_{k}$ and $\hat{F}_{k}$ are fermionic creation and anihilation operators in momentum space.
The matrix representation of the fermionic quadratic form is known as the the Bogoliubov de Gennes Hamiltonian

\begin{align}
          \label{eq:BdG}
\boldsymbol{H}_k&=\begin{bmatrix}
    2\hbar[g(t)-J(t)\cos k] & 2\hbar J(t)\sin k  \\
     2\hbar J(t)\sin k  & -2\hbar[g(t)-J(t)\cos k]  \\
         \end{bmatrix}
\end{align}
and describes a P-wave superconductor. Here the superconducting term describes the creation of pairs of fermions with opposite momenta~\cite{Kitaev2001}.

\section{Mapping QSP in the Heisenberg picture to the Schr\"odinger picture: The BCS ansatz\label{AppendixB}}
In the main text, we show that after applying Jordan-Wigner transformation and the discrete Fourier transform, we were able to reduce problem to a QSP sequence  in the Heisenberg picture using SU$(2)$ group. This was possible due to the pseudo-spin structure in momentum space. The natural question is how to map the QSP in terms of spins in real space.

A solution to this problem is to exploit the structure of the fermionic Hamiltonian Eq.~\eqref{eq:IsingHamitonianFermion} in the reciprocal space. This Hamiltonian breaks the conservation of particles and allows the creation of pairs of spinless fermions moving in opposite directions. The creation of pairs characterized by a time dependent pairing potential $\Delta_k(t)=2\hbar J(t)\sin k$ is odd under motion reversal symmetry $k\rightarrow -k$, which is a signature of a p-wave superconductor. As the excitations are created in pairs, one can show that any state of the system in the Schr\"odinger picture can be written using the well known BCS Ansatz from the theory of superconductivity~\cite{Dziarmaga2005}
\begin{align}
       \label{eq:BCSAnsatz}
              \ket{\Psi(t)}= \prod_{k>0}\left[v_k(t)+u_k(t)\hat{F}^{\dagger}_k\hat{F}^{\dagger}_{-k}\right] \ket{0}_k
    \ ,
\end{align}
where $\ket{0}_k$ is the vacuum for the $k$-th fermionic mode.
The key point of this approach is that the time-dependent coefficients appearing in the Ansatz can be obtained by using the relation
and has a general solution
\begin{align}
          \label{eq:FermionDyn}
\begin{bmatrix}
  u_{k} (t)\\
    v_{k}(t)
    \end{bmatrix}
= 
\begin{bmatrix}
    \mathcal{U}_k(t) &  \mathcal{V}_k^*(t)   \\
   \mathcal{V}_k(t)  & \mathcal{U}_k^{*}(t)  \\
    \end{bmatrix}
    .\begin{bmatrix}
    u_{k} (0)\\
    v_{k}(0)
    \end{bmatrix}
\ .
\end{align}
    The propagator in this equation is the same as the propagator $\boldsymbol{U}_k(t)$ in Eq.~\eqref{eq:FermionunitaryDyn} for the operators $\hat{F}_k(t)$ and $\hat{F}^{\dagger}_k(t)$ in the Heisenberg picture. One can think of this approach in terms of a pseudo spin approach, where the state of the two level system is described by a spinor $\psi^{\text{T}}_k(t)=[u_k(t),v_k(t)]$.
    
To have an intuitive understanding of this it is instructive to consider a simple example. Next we focus on the Ising Hamiltonian Eq.~\eqref{eq:IsingHamitonian} in the case of a constant transverse field $g(t)=g_0$ and in the absence of interactions $J(t)=0$. In this case the Bogoliubov de Gennes Hamiltonian Eq.~\eqref{eq:BdG} is diagonal $\boldsymbol{H}_k=2\hbar g_0\sigma_z$ and the propagator is $\boldsymbol{U}_k(t)=\exp{\left(-\mathrm{i}\boldsymbol{H}_k t/\hbar\right)}$. Now we can exploit the pseudospin picture to understand the physics of the problem. For example, when the states $\psi^{\text{T}}_k=(0,1)$ with negative energy $E^{(-)}_{k}=-2\hbar g_0$ are fully populated, we obtain the ground state of the system $\ket{\boldsymbol{0}}=\prod_{k>0}\ket{0}_k=\ket{\uparrow,\uparrow,\dots,\uparrow}$ with $\ket{\uparrow}$ and $\ket{\downarrow}$ being the eigenstates of $Z_j$. 
In the theory of the Ising model this is known as the paramagnetic ground state. In terms of fermions, this state describes a system with no pairs of counterpropagating excitations. The recipe to build up the excited states is to populate states with positive energies for a given wave vector $k_0$. That is, to create a pair of excitations with the desired momentum 
\begin{align}
          \label{eq:FermionDyn}
        \ket{1_{k_0},1_{-k_0}}&=\hat{F}^{\dagger}_{k_0}\hat{F}^{\dagger}_{-k_0}\ket{\boldsymbol{0}}=\mathrm{i}\sum_{i,j}e^{\mathrm{i}k_0(i-j)}\hat{f}^{\dagger}_i\hat{f}^{\dagger}_j\ket{\boldsymbol{0}}
        \nonumber \\
        &=\frac{\mathrm{i}}{2}\sum_{s,r}e^{\mathrm{i}k_0r}\hat{f}^{\dagger}_s\hat{f}^{\dagger}_{s+r}\ket{\uparrow,\uparrow,\dots,\uparrow}
        \nonumber \\
        &=\frac{\mathrm{i}}{2}\sum_{s,r}e^{\mathrm{i}k_0r}\ket{\uparrow,\uparrow,\downarrow_s,\uparrow\uparrow\dots\uparrow\uparrow,\downarrow_{s+r},\uparrow}
\ ,
\end{align}
where $\hat{f}^{\dagger}_s=(X_s+\mathrm{i}Y_s)/2\prod_{m=1}^{s-1}Z_m$ and $\hat{f}^{\dagger}_{s+r}=\left(\prod_{m=1}^{s+r-1}Z_m\right)(X_{s+r}+\mathrm{i}Y_{s+r})/2$. As $X_m^2=1$, we obtain the expression $\hat{f}^{\dagger}_s\hat{f}^{\dagger}_{s+r}=1/4(X_s+\mathrm{i}Y_s)\left(\prod_{m=s}^{s+r-1}Z_m\right)(X_{s+r}+\mathrm{i}Y_{s+r})$. The operator $\prod_{m=s}^{s+r-1}Z_m$ is the Pauli string connecting the sites $s$ and $s+r$.
To obtain this equation, we used the inverse Fourier transform $\hat{F}_k=\frac{e^{\mathrm{i}\frac{\pi}{4}}}{\sqrt{N}}\sum_j \hat{f}_{j} e^{-\mathrm{i}k j}$ to write the fermionic operators $\hat{F}^{\dagger}_{k_0}$ in terms of real space fermionic operators $\hat{f}^{\dagger}_j$. We also  inverted the Jordan- Wigner transformation Eq.~\eqref{eq:JordanWigner}  in order to write the fermionic operators $\hat{f}^{\dagger}_j$ in terms of spin operators in real space.
From the perspective of the pseudo spin, this is equivalent to apply a spin flip to the negative energy state with momentum $k_0$ to obtain a positive energy state $\psi^{\text{T}}_{k_0}=(1,0)$. In terms of the original spins in real space, this corresponds to the creation of a quantum superposition of localized spin flips.   
 
Alternatively, we can also study wave packets directly in the momentum representation. For example, for a two-particle initial state $\ket{\Psi(0)}=\sum_kG(k)\hat{F}^{\dagger}_{k}(0)\hat{F}^{\dagger}_{-k}(0)\ket{\boldsymbol{0}}$ with momentum distribution $G(k)$, the time evolution $\ket{\Psi(t)}=\sum_kG(k)\hat{F}^{\dagger}_{k}(t)\hat{F}^{\dagger}_{-k}(t)\ket{\boldsymbol{0}}$ can be obtained by considering the evolution of the operators in the Heisenberg picture
\begin{equation}
       \label{eq:EvolutionFermionic}
              \hat{F}^{\dagger}_{-k}(t)=\mathcal{V}_k(t)\hat{F}_{k}+\mathcal{U}^{*}_k(t)\hat{F}^{\dagger}_{-k}
       \ ,
\end{equation}
where $\mathcal{V}_k(t)$ and $\mathcal{U}^{*}_k(t)$ are matrix elements of the propagator $\boldsymbol{U}_k(t)$ in Eq.~\eqref{eq:FermionunitaryDyn} for the operators $\hat{F}_k(t)$ and $\hat{F}^{\dagger}_k(t)$ in the Heisenberg picture. Thus, the time evolution of the wave packet can be written as
\begin{align}
       \label{eq:EvolutionWavePacket}
              \ket{\Psi(t)}&=\sum_kG(k)\mathcal{V}_{-k}(t)\mathcal{U}^{*}_{k}(t)\ket{\boldsymbol{0}}
              \nonumber \\&+
              \sum_kG(k)\mathcal{U}^{*}_{k}(t)\mathcal{U}^{*}_{-k}(t)\ket{1_{k},1_{-k}}
       \ .
\end{align}
Importantly, this wave packet can be interpreted as a quantum superposition of the paramagnetic ground state and a wavepacket of two spin flip excitations by considering Eq.~\eqref{eq:FermionDyn}.

These simple examples captures the essence of our approach. To design a QSP sequence using the generators of the Onsager algebra in the Schr\"dinger picture is a cumbersome task. However, we can easily design a QSP sequence in the Heisenberg picture for the operators $\hat{F}_k$ and $\hat{F}_k^{\dagger}$ using the SU$(2)$ pseudo spin representation. In turn, QSP sequences giving the propagator $\boldsymbol{U}_k(t)$ in the pseudo-spin representation can be directly mapped to operations in real space using the BCS Ansatz in Eq.~\eqref{eq:BCSAnsatz}.

\section{QSP Sequences for general $\theta$ \label{AppendixC}}
In this appendix, we discuss QSP sequences for general values of $k$ and an unknown $\theta$. As we are processing two independent variables, the QSP sequence is more complicated that the one discussed in the main text. In our manuscript, one of the restrictions we found is that the signal and signal processing operations are rotations along non-orthogonal axes. To overcome this restriction, we can define a modified QSP sequence for the Onsager algebra
\begin{align}
          \label{eq:GeneralIsingQSP}
\hat{U}^{M}_{\vec{\phi}}(\theta)=\prod^d_{r=1}e^{\mathrm{i} \theta\sum^{N}_{j=1}X_j X_{j+1}}e^{\mathrm{i}\frac{\pi}{4}\sum^{N}_{j=1} Z_j}e^{-\mathrm{i} \theta\sum^{N}_{j=1}X_j X_{j+1}}
e^{\mathrm{i}\phi_{r}\sum^{N}_{j=1} Z_j}
\ .
\end{align}
In arrays of superconducting qubits, if the parameter $\theta$ is known, its sign can be controlled using microwave control lines~\cite{mi2022time}. When the parameter $\theta$ is unkown, its sign can be effectively changed from positive to negative by applying $\pi/2$ rotations along the $Z$ axis to the even or odd sites.
Next, let us explore the form of our modified QSP sequence in momentum space, which reads
\begin{align}     
 \label{eq:GeneralBogIsingQSP}
\boldsymbol{U}^{M}_{k,\vec{\phi}}(\theta)&=
\prod^d_{r=1}e^{2\mathrm{i} \theta (\sigma_z \cos k-\sigma_x \sin k)}e^{-2\mathrm{i} \theta (\sigma_z \cos k+\sigma_x \sin k)}
e^{-2\mathrm{i}(\phi_{r}+\pi/4)\sigma_z}
\ .
\end{align}
It is worth noting that the rotation $e^{\mathrm{i}\frac{\pi}{4}\sum^{N}_{j=1} Z_j}$ maps to a pseudo spin rotation $e^{-\mathrm{i}\frac{\pi}{2}\sigma_z}$ in momentum space. Also, the first two terms in the QSP sequence are rotations along an the axis $\hat{n}_k=[-\sin k,0, \cos k]$ and its reflection $\hat{m}_k=[-\sin k,0, -\cos k]$ along the x axis. By using the fundamental properties of SU$(2)$ rotations, we obtain the general QSP sequence
\begin{align}     
 \label{eq:GeneralOrthogonalQSP}
\boldsymbol{U}^{M}_{k,\vec{\phi}}(\theta)&=
\prod^d_{r=1}e^{\mathrm{i} \Omega_k (A_k\sigma_x +B_k\sigma_y )}
e^{-2\mathrm{i}(\phi_{r}+\pi/4)\sigma_z}
\ ,
\end{align}
where $\cos\Omega_k=\cos^2(2\theta)+\cos (2k)\sin^2(2\theta)$ and the new axis is defined by the parameters
\begin{align}     
 \label{eq:AxisParameters}
A_k&=-\frac{\sin k \sin (4\theta)}{\sin\Omega_k}
\nonumber \\
B_k&=\frac{\sin(2k) \sin^2 (2\theta)}{\sin\Omega_k}
\ .
\end{align}
Even if the parameter $\theta$ is unknown, this QSP sequence is composed by rotations along orthogonal axis. However, in contrast to the QSP sequence discussed in the main text, here the signals parameters $k$ and $\theta$ define the rotation axis in the $x-y$ plane in a nonlinear fashion, while the signal processing takes place along the $z$ axis.

\section{Space-time Dual QSP for $k=\pi/2$ \label{AppendixD}}
In this appendix, we discuss the QSP sequence $\boldsymbol{V}_{k,\vec{\Phi}}$ for the space-time dual quantum circuit in the main text. As a first step, it is useful to consider the QSP sequence in momentum space
\begin{align}
\label{eq:ThetapiNew}
\boldsymbol{V}_{k,\vec{\Phi}}= e^{i\pi/4 \sigma_z }\left(\prod_{r=1}^{d}e^{-ik \sigma_x}  e^{-\mathrm{i}\frac{\pi}{2}\left(1 -4\epsilon  \right)\sigma_z}\right)e^{-i\pi/4 \sigma_z },
\end{align}
where we took $\phi_r = \pi/2 (1-2\epsilon)$ in the definition of $\vec{\Phi}$ according to Eq. ~\eqref{eq:QSVT_w_fixed_angle}. 

We notice that the evolution $e^{2\mathrm{i} \theta (\sigma_z \cos k-\sigma_x \sin k)}$ in Eq.~\eqref{eq:Xevolve} of the fermionic operators under the Ising interaction  
becomes $e^{\mp\mathrm{i} \pi/2 \sigma_x }=\mp i \sigma_x$ when $k=\pm \pi/2$ and $\theta=\pi/4$. Then, we can write the composite pulse sequence (up to a constant phase) as 
\begin{align}
\nonumber \boldsymbol{V}_{\mp\pi/2,\vec{\Phi}} &= (\mp i)^{d}\sigma_x e^{-i\pi(1-2\epsilon)\sigma_z}\sigma_x e^{-i\pi(1-2\epsilon)\sigma_z} \cdots \sigma_x e^{-i\pi(1-2\epsilon)\sigma_z}\\
&\propto \bigg\{ \begin{array}{cc} \mathbf{I}& \mathrm{if} \, d \in \mathrm{even} \\\sigma_x e^{-i\pi(1-2\epsilon)\sigma_z} & \mathrm{if} \, d \in \mathrm{odd}
\label{eq:Ukpmpi}
\end{array} 
\end{align}
Hence, the resulting unitary approximates the dynamics up to an error $\epsilon$ in the phase rotation.


%

\end{document}